\documentstyle[psfig,graphicx]{mn}

\def\gapprox{\lower.4ex\hbox{$\;\buildrel >\over{\scriptstyle\sim}\;$}}
\def\lapprox{\lower.4ex\hbox{$\;\buildrel <\over{\scriptstyle\sim}\;$}}

\title{Single pulses as emission from multiple subsources}

\author[Luo]
      {Qinghuan Luo\\
        School of Physics, The University of Sydney, NSW 2006, Australia\\
}

\date{
          --- Received
         in original form February, 2004
        }

\pubyear{2004}

\begin{document}

\maketitle

\begin{abstract}
Single pulses of pulsar radio emission are modeled as superposition of radiation 
originating from many small subsources that are randomly distributed in the emission
region. The individual subsources are given an intrinsic finite angular beaming. This model can
produce fluctuations in intensity and pulse profiles that are similar to the microstructure 
observed in some pulsars. Statistics of the phase resolved flux density of a simulated 
single pulse can be approximated by a lognormal distribution, which is in good 
agreement with observations.
\end{abstract}


\section{Introduction}

Observations of single radio pulses from pulsars show a high degree of polarization and rapid 
variation in intensity, often with spiky structures called the microstructure (Craft, 
Comella \& Drake 1968; Hankins 1971).  Microstructures seem to occur simultaneously 
over a wide range of frequencies (e.g. Rickett, Hankins \& Cordes 1975),  
suggesting that they are closely associated with intrinsic properties at the emission origin. 
Three types of mechanisms have been 
considered as possible causes for the microstructure: (1) temporal modulation of the 
relevant emission process or wave propagation, e.g. fluctuations in radiation as the result of modulational 
instability associated with the coherent emission process (Harding \& Tademaru 1981;
Chian \& Kennel 1983; Weatherall 1998) or as the result of modulation of radio waves due to a 
low frequency wave (Machabeli et al 2001); (2) beaming effect  attributed to individual emission beams 
sweeping across the line of sight (Hankins 1972); (3) the combined effect of relativistic beaming and
a nonuniform distribution of the emitting plasma along the curved field lines (Luo \& Melrose 2004). 
The broadband nature of microstructures (Lange et al. 1998; Kramer, Johnston \& van Straten 
2002) appears to favor the second or third mechanisms, which are
intrinsic to the emission origin. The third one differs from the first in that the mechanism 
requires inhomogeneities in a longitudinal distribution of the emitting plasma along the curved 
field lines. Since nonuniform structures produce fluctuations within the individual emission 
beam, the third mechanism can lead to much narrower microstructures than predicted by 
the conventional beaming model. 

In both beaming models (mechanisms (2) and (3)) for microstructures a nonuniform, nonstationary 
pair cascade above the polar cap (PC) is required. In the widely-dicussed PC models the 
radio emission is generally assumed to be produced in the outflowing electron/positron pair 
plasma within the pulsar magnetosphere (e.g. Melrose 2000). 
A rotation-induced electric field above the PC accelerates 
primary electrons (or positrons) to ultrarelativistic energies; these particles radiate high 
energy $\gamma$-rays, which decay into pairs in the strong pulsar magnetic field, forming an 
outflowing pair plasma (Sturrock 1971; Arons \& Scharlemann 1979; Daugherty \& Harding 1982). 
The assumption of steady pair production, which is a major feature of the current models,
may be oversimplified since particle acceleration and the subsequenct pair cascade are strongly
affected by an external current flow (e.g. Lyubarskii 1992; Shibata 1997; Mestel 1998) and 
the pair cascade can be nonstationary and occur nonuniformly across the PC. Therefore, it is 
plausible that the outflowing pair plasma is highly inhomogeneous in space and varies rapidly in time. 
Owing to the beaming, a radiation pattern generated in such inhomogeneous, nonstationary 
plasma that streams relativistically along the curved magnetic 
field lines, must have nonuniform transverse structures that sweep across the line of sight.
This can give rise to both the fluctuations and the pulse structure.

In this paper sinlge pulse emission is modeled as radiation from many subsources randomly
distributed in the emission region. The fluctuations and pulse structure caused by the random 
distribution are considered. Individual subsources radiate in the field line direction with 
a finite angular spread. The observed single pulse is considered as superposition of 
many subsources that naturally provide the microstructure. Their random distribution
leads to fluctuations in intensity from pulse to pulse. To model emission from multiple 
sources, one assumes that the radio emisison is in the plasma natural modes that 
can escape the pulsar magnetosphere (e.g. Melrose 2000). The radio emission propagates away 
from the emission region with the polarization being that of the natural modes up to the 
polarization limiting region (PLR), beyond which the polarization is no longer affected by 
the plasma and is frozen to its value at PLR (Melrose \& Stoneham 1977; Barnard \& Arons 1986). 
Specifically, single pulses are simulated numerically with the polarization properties derived from 
the local plasma dispersion at the PLR. 

The multiple subsource model is described in details in Sec.~2. Numerical simulation of single 
pulses and the implications for the interpretation of the microstructure and fluctuations in intensity
are discussed in Sec.~3. Conclusions are given in Sec.~4.

\section{Radiation from randomly distributed subsources}

Single pulse emission is modeled as superposition of emission from a random distribution of
subsources in the emission region. A nonstationary pair cascade above the PC produces a 
nonsteady, inhomogeneous pulsar plasma. The corresponding source is then highly inhomogeneous 
and nonstationary. One may model such source in terms of a distribution of multiple subsources. 
Polarized radio emission can be completely described by the Stokes parameters ($I$, $U$, $Q$, $V$).
Assuming that these subsources are not phase related, the Stokes parameters can be written as a sum of those 
from individual emitters,
\begin{eqnarray}
I&=&\sum_i(I^+_i+I^-_i), \nonumber\\
U&=&\sum_i\xi^l_i(I^+_i-I^-_i)\sin2\chi_{i},\nonumber\\
Q&=&\sum_i\xi^l_i(I^+_i-I^-_i)\cos2\chi_{i},\nonumber\\
V&=&\sum_i\xi^c_i(I^+_i-I^-_i),
\label{eq:stokes}
\end{eqnarray}
where $I^\pm_i$ are the intensities of radiation from the $i$th source in two orthogonal modes 
$\pm$, $\chi_i$ is the position angle (PA) of a ray originating from the $i$th subsource
in the observer's direction, and $\xi^l$ and $\xi^c$ are the degree of linear and circular polarization,
given by
\begin{equation}
\xi^l={1-T^2\over1+T^2},\ \ \ \
\xi^c={2T\over 1+T^2},
\label{eq:xi}
\end{equation}
with $T$ the polarization ellipticity of the $+$ mode. 

\subsection{Relativistic beaming}

\begin{figure}
\centerline{\psfig{file=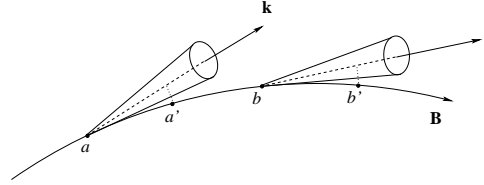,width=7cm}}
\caption{Radio emission from multiple subsources. The radio waves are emitted from the subsources 
at $a$ and $b$, which are beamed in a narrow cone with a half opening angle of $1/\Gamma_s$ 
in the field line direction.  The polarization is assumed to be decoupled from the local 
plasma and field line curvature at $a'$ and $b'$.}
\label{fig:ray}
\end{figure}

In the observer's frame, radiation from subsources is beamed in the direction of motion of 
the source with an angular spread $\Delta\Omega_0=1/\Gamma_s$, where $\Gamma_s$ 
is the bulk Lorentz factor of the source (as shown in Figure~\ref{fig:ray}). Both aberration and 
refraction can change the beaming direction substantially (Blaskiewicz, Cordes \& Wasserman 1991; 
Petrova 2000; Fussell \& Luo 2004). The latter effect can also cause the two modes to 
separate (Melrose \& Stoneham 1977). All these effects are ignored in the following discussion.
The angular distribution of the intensity of the individual subsource is written as
\begin{equation}
I^\pm_i(\hat{\bf k})=I^\pm_{0i}
\exp\left[-{2(1-\cos\theta_{kb})\over\Delta\Omega^2_0}\right],
\label{eq:beaming}
\end{equation}
where $\hat{\bf k}$ is the direction of the line of sight, $I^\pm_{0i}$ is the central intensity 
of the beam and $\theta_{kb}$ is the propagation angle with respect to the field line direction
$\hat{\bf b}$. 

One may express $\theta_{kb}$ in terms of the polar angles of $\hat{\bf k}$ and $\hat{\bf b}$ given
respectively by $(\theta_k,\phi_k)$ and $(\theta_b,\phi_b)$. In the relativistic limit $\Gamma_s\gg1$,
it is convenient to use the small angle approximation $|\theta_k-\theta_b|\ll1$ and
$|\phi_k-\phi_b|\ll1$. Then, the propagation angle in the observer's frame is written as  
$\theta_{kb}\approx \left[(\theta_k-\theta_b)^2+
\sin\theta_k\sin\theta_b(\phi_k-\phi_b)^2\right]^{1/2}$.  

\subsection{Distribution of subsources}

The distribution of subsources is closely related to that of the pair cascade
above the PC. The dominant pair creation process is single photon decay in 
the pulsar magnetic field, which is the most efficient on the field lines with the
smallest radius of curvature. Therefore, pair production should be peaked on 
the field lines near the surface boundary subtended by the last open field lines where
the electric field is strong and the radius of field line curvature is small (e.g. 
Arons \& Scharlemann 1979; Daugherty \& Harding 1982). In contrast, there are
few pairs created near the magnetic pole where the field line curvature tends to
become infinite large, or on the last open field lines on which the parallel electric field 
starts to drop off rapidly to become zero (on the closed field lines).

Similar to the distribution of pair production, one assumes that subsources have a 
distribution peaked near the last open field lines.
Since the pulsar plasma flows relativistically along open field lines, one
may assign each subsource a magnetic polar coordinate $(r,\theta,\phi)$, where 
$(\theta,\phi)$ are the polar angles of the field lines on the PC, and $r$ 
the radial distance of the source to the star's center. One assumes that the distribution 
is peaked on the field lines $\theta_c\equiv\epsilon_c\theta_d$, where $0<\epsilon_c<1$ is
the parameter that characterises how close the peak is to the outer rim of the PC where
the last open field lines intercept the stellar surface. The PC is defined by the
half-opening angle $\theta_d=(\Omega R_0/c)^{1/2}$, where $\Omega=2\pi/P$, $R_0=10^6\,\rm cm$
is the star's radius. The general form of the distribution is then given by
\begin{equation}
f(\theta,\phi)=g(\phi)\exp\left[-{\textstyle{1\over2}}\left({\theta-
\theta_c\over\epsilon\theta_c}\right)^{2p}\right],
\label{eq:distribution}
\end{equation}
where $0\leq\theta\leq\theta_d$ and $0\leq\phi\leq2\pi$,
$\epsilon=\{\epsilon_1,\epsilon_2\}$ and $p>1$ are three parameters that characterise the
general shape of the distribution centered at $\epsilon_c\theta_d$, with $\epsilon_1$ characterising
the slope at $\theta<\theta_c$ and $\epsilon_2$ at $\theta>\theta_c$. One has a
Gaussian distribution in $\theta$ for $p=1$. Since pair creation is generally not axially 
symmetric with respect to the magnetic pole, one includes the azimuthal dependence  $g(\phi)$
in (\ref{eq:distribution}).

\subsection{Polarization}

The polarization is assumed to follow that of the plasma natural modes until the PLR beyond which
the wave retains its polarization (Melrose \& Stoneham 1977; Melrose \& Luo 2004a). Thus, the 
polarization in (1) is evaluated at the PLR. Specifically, the polarization ellipse $T$ 
and PA $\chi_i$ are written as a function of the distance $d\leq R_{LC}=c/\Omega$ along the 
propagation path from the source to the PLR, called the decoupling distance, where $R_{LC}$ is 
the light cylinder radius. 
Similar to the conventional rotating vector model (Radhakrishnan \& Cooke 1969), the PA at the 
PLR can be determined by the orientation of the local field line plane (at the PLR). Since we 
emphasize the microstructures and fluctuations in intensity, to simplify numerical calculation the 
ellipticity $T$ is evaluated in the cold electron-positron plasma with a net charge density
$\eta=(n_+-n_-)/(n_++n_-)\neq0$, where $n_\pm$ are the number density of electrons (positrons).
A detailed discussion of the ellipticity including the effect of the relativistic distribuion is 
given in Melrose \& Luo (2004a, b). For $\omega\gg\omega_p$, where $\omega_p$ is the
plasma frequency, the natural modes can be regarded
approximately as transverse and $T$ is approximately a Lorentz invariant. It is then convenient
to calculate $T$ in the plasma rest frame (Melrose \& Luo 2004a).

In the low frequency approximation, $\omega\ll\Omega_e$, $T$ changes its sign if the propagation 
angle in the plasma rest frame sweeps across $90^\circ$, which is Lorentz transformed to
$\theta_{kb}\sim1/\Gamma_s$ in the observer's frame (Melrose \& Luo 2004a). The 
ellipticity can also change sign if the net charge density changes its sign, e.g. it may occur
on null surface where the magnetic field direction is perpendicular to the rotation axis
(Figure 2d and 2f). 

\begin{figure*}
\begin{center}
\psfig{file=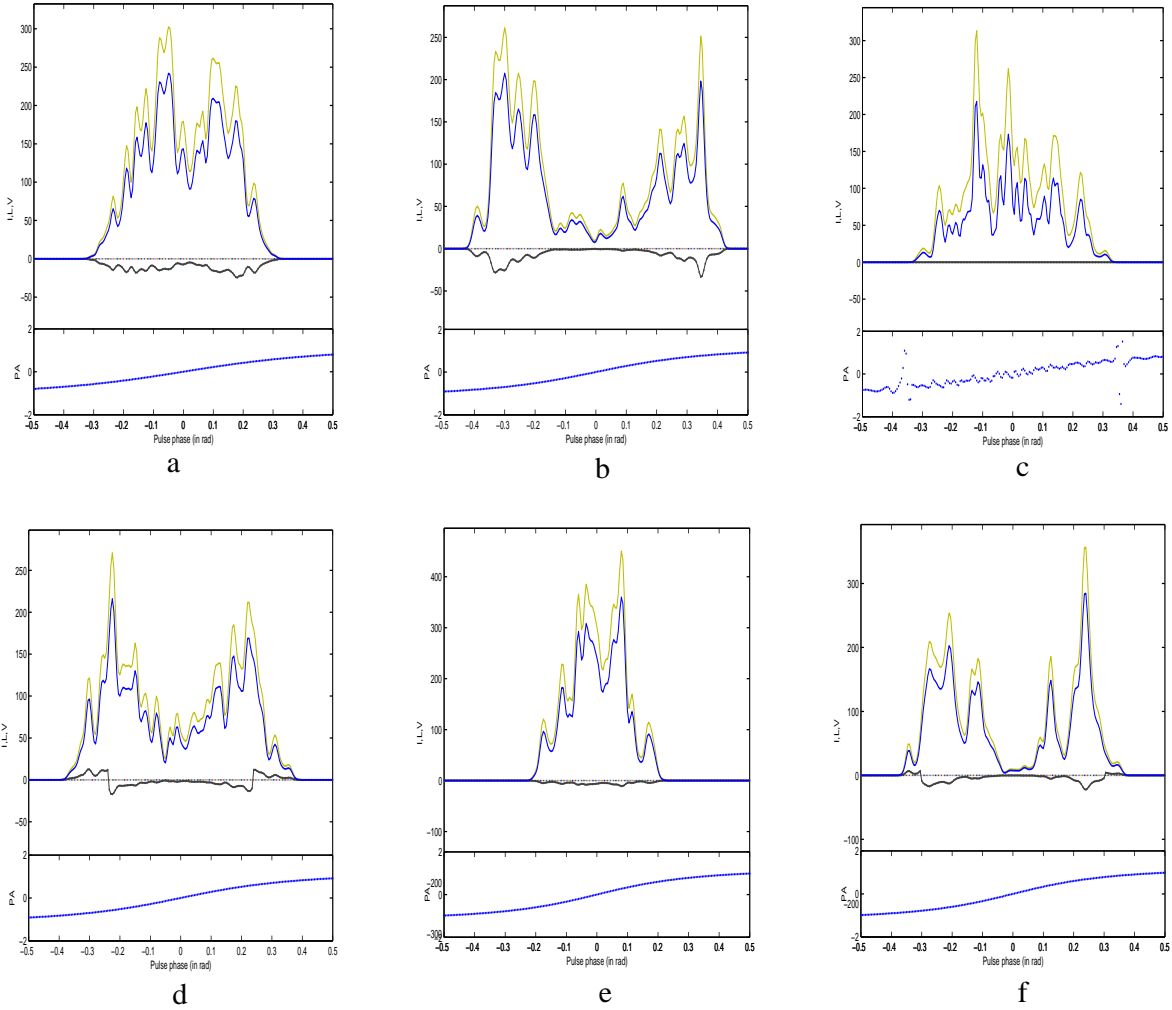,width=16cm}
\end{center}
\caption{Simulated single pulses for a pulsar with $P=0.1$, $\alpha=\pi/3$. 
The three (light, solid and dotted) plots in each case correspond respectively
to the three intensities $I$, $L=(U^2+Q^2)^{1/2}$, $V$. The corresppnding PA variation 
is shown below each profile. a: $\alpha_m=0.9$, $d=200$, the viewing angle (relative to 
the rotation axis) $i=\pi/3+0.4$, and the subsources distributed within the radial range 
$r=50-50.5$; b: As in a, but with
$d=250$, $i=\pi/3+0.3$; c: As in a, but with $d=5$; d: $\alpha_m=0.9$, $i=\pi/3+0.3$,
$r=42-42.5$, $d=250$; e: $d=280$, $i=\pi/3+0.25$, $r=20-20.5$; f: As in e, but with $r=35-35.5$. 
The bulk Lorentz factor of the emitting plasma is $\Gamma_s=100$.
The subsources are distributed within the colatitude range $0\leq\theta_*\leq\theta_d$ 
centered at $\theta_c=0.8\theta_d$, with a width $\Delta\theta_c=0.1\theta_d$. }
\label{fig:profile}
\end{figure*}

\section{The result and discussion}

In the simulations one draws a random distribution of subsources from 
(\ref{eq:distribution}) with $g(\phi)=1$, a radial range of $r$ to $r+\Delta r$ and angular ranges of
$0\leq\theta\leq\theta_d$ and $0\leq\phi\leq 2\pi$ to evaluate the Stokes 
parameters (\ref{eq:stokes}). The simulated pulse profiles
are shown in Figure \ref{fig:profile} for a pulsar with the inclination $\alpha=\pi/3$, 
the pulse period $P=0.1\, \rm s$. The pulsar magnetic
field is assumed to be a dipole with $B=10^{12}\, \rm G$ on the PC.
For the efficiency of the numerical calculation, one considers a small radial range 
$\Delta r\ll r$, where all radial distances are in units of the star's radius $R_0$. 
The plasma has a bulk Lorentz factor $\Gamma_s=100$ with a net charge
density $\eta=-0.1$. A case with sign change, i.e. $\eta=-0.1$ for 
$\hat{\bf \Omega}\cdot\hat{\bf b}<0$ and $\eta=0.1$ for $\hat{\bf \Omega}\cdot\hat{\bf b}>0$, 
where $\hat{\bf \Omega}$ is the spin axis, is also conisdered.  One further
assumes the emission frequency $\omega/\omega_p=20$ in the pulsar frame and that all subsources 
have the same central 
intensity $I_{0i}={\rm const}$. The distribution of subsources is characterised by $p=1$ (Gaussian), 
$\epsilon_c=0.8$ and $\epsilon_1=\epsilon_2=0.1/\epsilon_c$. Because the 
distribution of subsources along the line of sight is random, the total intensity as well 
as the polarized intensities fluctuate across the pulse phase.

\subsection{One dominating mode}

As an example one considers the case of one dominating mode with $\alpha_m\equiv I^+/(I^++I^-)=0.9$. 
All six simulated pulses in Figure 2 show depolarization in intensity, though the individual 
subsources are 100\% polarized. One may estimate the polarized intensity of the emergent radiation 
as $I_p\leq (2\alpha_m-1)I\approx0.8I$ with $I=I^++I^-$. The depolarization is due to both the 
effects of mode mixing and cancellation of polarization of radiation from sources originating
from different divergent field lines. A spread in the radiation beam of the individual subsource
also leads to depolarization. The latter two effects are especially important if the decoupling distance 
$d$ is short and the line of sight samples a large number of subsources originating from the field lines 
that diverge away from the viewing direction (cf. Figure 2c).

PA variation can be distorted because of the cancellation effect on the polarization 
within the emission beam, which is strong
at the emission origin. An example of scattering in PA is shown Figure 2c. As one chooses
a small $d$, due to the strong cancellation effect on the polarization, 
one has broad scattering in PA. Some pulsars are seen to have a large spead in PA (e.g. 
Gil \& Lyne 1995; Karastergiou et al. 2002), which can be explained by the combination of a short 
decoupling distance $d$ and a wider angular beaming of the subsource.

The single pulses in Figure 2 have a relatively large CP as the parameters used in the
simulation correspond to the regime of the aberrated backward circular polarization (ABCP)
(Melrose \& Luo 2004a). In this regime, waves propagate backward in the local plasma rest 
frame, and are elliptically polarized.

\subsection{Microstructures}

The simulated pulses show substructures that are similar to the microstructrue seen in 
observations. The profiles shown in Figure~\ref{fig:profile} are the
snapshots of the radiation pattern produced from a random distribution of subsources.
The pulse structures are due to the transverse (across the field lines) effect
only. Because of the relativistic beaming each subsource has a natural angular width of 
$1/\Gamma_s$, which gives rise to a typical time scale given by (Cordes 1979)
\begin{equation}
\tau_\mu\approx {P\over2\pi \Gamma_s}.
\label{eq:micro1}
\end{equation}
The recent study of cascade above the PC confirms that the secondary pairs have a broad distribution 
peaked at a moderate Lorentz factor around $\Gamma_s\approx 10^2$, which is not particularly sensitive
to the pulse period (Zhang \& Harding 2000; Hibschman \& Arons 2001; Arendt \& Eilek 2002). 
Hence, (\ref{eq:micro1}) predicts a approximately linear relation with $P$. 
Observations of microstructures seem consistent with this linear relation
(Kramer, Johnston \& van Straten 2002). For $\Gamma_s=100$, $P=0.1\, \rm s$, one has 
$\tau_\mu=159\,\mu\rm s$. 

The pulse structure can be smeared out substantially if the subsources are densely packed 
in the emission region. This may explain why some pulsars do not show any microstructure.

Observations appear to suggest that microstructures may have a much narrower width
corresponding to a time scale much shorter than (\ref{eq:micro1}) (e.g. Popov et al 2002). However,
nanosecond structures have been clearly confirmed only for giant pulses (Hankins et al 2003), which
are thought to have a different origin from normal pulses (Romani \& Johnston 2001). Such 
rapid variation can be produced by rapid temporal modulation such as plasma turbulence
(Hankins et al 2003) or the combined effect of the field line curvature and an inhomogeneous 
distribution of sources along the field lines (e.g. Gil 1985; Luo \& Melrose 2004).  In the latter 
case, due to the relativistic beaming and the field line curvature, the observer can only see radiation 
within a very small temporal window. For a source moving along the curved field lines, with a 
longitudinal extent $\Delta L_e< 2R_c/\Gamma_s$, where  $R_c$ is the radius of curvature, an 
observer can only see the radiation within the time interval given by (Luo \& Melrose 2004)
\begin{equation}
\tau_\mu={1\over\Gamma^3_s\omega_R},
\label{eq:micro2}
\end{equation}
where $\omega_R=c/R_c$. If the source is nonuniform along the field lines over the length scale less 
than $2R_c/\Gamma_s$, the radiation must have temporal structures with the time scale  
given by (\ref{eq:micro2}). Simulation of such rapid variation in intensity can be done 
in the same way as that leads to Figure 2 except that one draws different random distributions of 
subsources within one pulse period.

\subsection{Fluctuations in intensity}

\begin{figure}
\psfig{file=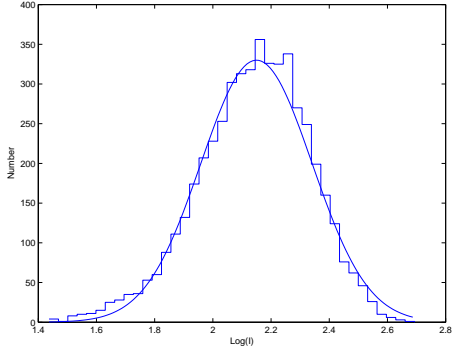,width=6cm}
\caption{Distribution of single pulse flux intensities (the histogram plot).
The solid line corresponds to a fit to the lognormal distribution. The parameters are as 
in Figure \ref{fig:profile}a.}
\label{fig:lognorm1}
\end{figure}

Since the number of subsources beaming into the line of sight is random, the relevant intensity
varies from pulse to pulse. Relative intensities of a simulated pulse can be obtained at a 
particular pulse phase by repeating the simulation a large number of times. The distribution 
of the intensities at the pulse phase $\phi=0.1\, \rm rad$ near the pulse peak is shown 
in Figure~\ref{fig:lognorm1}.  The distribution can be 
reasonably well fitted by a lognormal distribution, which appears to be in good agreement 
with observations of single pulses (e.g. Cairns, Johnston \& Das 2001; Kramer,
Johnston \& van Straten 2002). In this model, one expects the simulated fluctuating intensity
to follow roughly a lognormal distribution if the dominant contribution to the 
total intensity is from those subsources with beaming at an angle, 
$\Delta\Theta_i=(1-2\cos\theta_{kb})\ll\Delta\Omega_0$, with respect to the line of 
sight. Since the intensity from the contributing subsource is 
given by (\ref{eq:beaming}), the total intensity can be written as
\begin{eqnarray}
I&=& \sum_iI_{0i}\exp\left(-{\Delta\Theta^2_i\over\Delta\Omega^2_0}\right)
\nonumber\\
&\approx& I_0\exp\left(-{1\over I_0}\sum_iI_{0i}{\Delta\Theta^2_i\over\Delta\Omega^2_0}
\right),
\label{eq:Iexp}
\end{eqnarray} 
where $I_0=\sum_i I_{0i}$. Assuming a large number of such subsources, according to 
the central limit theory (CLT) one expects ${\rm log}I$ to approach a lognormal 
distribution.

The distribution may deviate from lognormal if the total intensity results 
predominantly from those subsources with $\Delta\Theta_i>\Delta\Omega_0$, which 
may occur at the edge of the pulse profile.

\section{Conclusions}

We consider a model for single pulses, based on the hypothesis that the radiation
is superposition of many subsources randomly distributed in the open field line region.
The subsources radiate within a finite beaming angular range $\sim 1/\Gamma_s$. 
Numerical simulations are carried out to model the pulse profile. The model can predict
the basic features of microstructures and fluctuations in intensity. Simulated single pulses show 
substructures due to individual emission beams from subsources that are randomly distributed
along the line of sight. In this mulitple subsource model, one simulates fluctuations in intensity by 
repeating the simulation of a random distribution of subsources. The phase resolved intensities near 
the peak of the simulated single pulse are well fitted by a 
lognormal distribution. Other effects such as wave absorption/amplification (e.g.
Cairns, Johnston \& Das 2001), scattering in an inhomogeneous intervening plasma (Ishimaru 1997)
can contribute significantly or even dominantly to the total fluctuations. However, to include all 
these effects in the model one requires specific models for these processes in the pulsar magnetosphere. 

The model predicts a large scattering in PA if the PLR is close to the emission origin
(i.e. the decouple distance is short) or the beaming is wide.
The spread in PA is due to the spread in angles between the lin eof sight and the range of diverging field lines
from which radiation can be seen at any given time. The model reproduces ABCP proposed by Melrose \& Luo (2004a). 
The polarization is modeled in a way different from the conventional rotating vector
model (e.g. Radhakrishnan \& Cooke 1969; Blaskiewicz, Cordes \& Wasserman 1991)
in that the polarization is elliptical and that it is determined at the PLR, not at the emission point.
The PLR is characterised by its distance $d$ to the emission origin and in this model $d$ is treated
as a free parameter but constrained by $d<R_{LC}$. Further work on models in which $d$ can be determined
quantitatively is needed.

\section*{Acknowledgement}

The author thanks Don Melrose and Jean-Pierre Macquart for useful discussion and comments.


\begin{thebibliography}{22}

\bibitem{arendt-eilek02}
Arendt  P.N.Jr,  Eilek  J.A., 2002, ApJ, 581, 451

\bibitem{arons-scharlemann79}
Arons J., Scharlemann E.T., 1979, ApJ, 231, 854.


\bibitem{barnard-arons86}
Barnard J.J.,  Arons J. 1986, ApJ, 302, 138


\bibitem{blaskiewicz-etal91}
Blaskiewicz M., Cordes J.M., Wasserman I. 1991, ApJ, 370, 643

\bibitem{cjd01}
Cairns I.H., Johnston S., Das P. 2001, ApJ, 563, L65

\bibitem{cr77}
Cheng A.F., Ruderman M.A. 1977, ApJ, 212, 8000

\bibitem{ck83}
Chian A.C., Kennel C.F.  1983, ApSS, 97, 9

\bibitem{c79}
Cordes J.M. 1979, Aust. J. Phys. 32, 9

\bibitem{cordes-etal78}
Cordes J.M., Rankin J.M., Backer D.C. 1978, ApJ, 223, 961

\bibitem{ccd68}
Craft H.D., Comella J.M., Drake F. 1968, Nat, 218, 1122

\bibitem{daugherty-harding82}
Daugherty J.K., Harding A.K., 1982, ApJ, 252, 337

\bibitem{fl04}
Fussell D., Luo Q. 2004, MNRAS, in press

\bibitem{g85}
Gil J. 1985, ApSS, 110, 293

\bibitem{gil-lyne95}
Gil J., Lyne A. 1995, MNRAS, L55


\bibitem{h71}
Hankins T.H. 1971, ApJ, 169, 487

\bibitem{h72}
Hankins T.H. 1972, ApJ, 169, 487

\bibitem{h03}
Hankins T., Kern J.S., Weatherall J.C., Eilek J.A. 2003, Nat, 422, 141

\bibitem{ht81}
Harding A., Tademaru E. 1981, ApJ, 243, 597.

\bibitem{hibschman-arons01}
Hibschman J., Arons J. 2001, ApJ, 546, 382

\bibitem{i97}
Ishimaru A. 1997, Wave Propagation and Scattering in Random Media, (IEEE Press)

\bibitem{k-etal02}
Karastergiou A., Kramer M., Johnston S., Lyne A.G., Bhat N.D.R., 
Gupta Y. 2002, A\&A, 391, 247

\bibitem{Kramer-etal02}
Kramer M., Johnston S., van Straten W. 2002, MNRAS, 334, 523

\bibitem{letal98}
Lange C., Kramer M., Wielebinski R., Jessner A. 1998, A\&A, 332, 111

\bibitem{luo-etal02}
Luo Q., Melrose D.B., Fussell D. 2002, Phys. Rev. E66, 026405

\bibitem{lm03}
Luo Q., Melrose D.B. 2004, in Young Neutron Stars and Their Environments, IAU218, in press

\bibitem{lm88}
Lyne A., Manchester R.N., 1988, MNRAS, 234, 477

\bibitem{lp92}
Lyubarskii Yu. E. 1992, A\&A, 261, 544

\bibitem{metal01}
Machabeli G.Z., Khechinashvili D., Melikidze G. \& Shapakidze D. 2001, MNRAS, 327, 984

\bibitem{ms00}
McKinnon M. M., Stinebring D. D., 2000, ApJ, 529, 433


\bibitem{m00}
Melrose D. B., 2000, in Kramer M., Wex N., Wielebinski R., eds, ASP
Conf. Ser. Vol. 202, Pulsar astronomy - 2000 and beyond, Astron. Soc.
Pac., San Francisco, p. 721

\bibitem{ml03a}
Melrose D.B., Luo Q., 2004a, MNRAS, in press

\bibitem{ml03b}
Melrose D.B., Luo, Q., 2004b, PRE, submitted 

\bibitem{ms77}
Melrose D.B., Stoneham R.J. 1977, Proc. ASA, 3, 120

\bibitem{m98}
Mestel, L. 1998, Stellar Magnetism, (Oxford University Press)

\bibitem{p00}
Petrova S. 2000, A\&A, 360, 592

\bibitem{p-etal02}
Popov M.V., Bartel N., Cannon W.H., Novikov Yu.A.,
Kondratiev V.I., Altunin V.I. 2002, A\&A, 396, 171

\bibitem{radhakrishanan-cooke69}
Radhakrishnan V., Cooke D.J. 1969, Ap. Letters, 3, 225

\bibitem{rhc75}
Rickett B.J., Hankins T.H., Cordes J.M. 1975, ApJ, 201, 425

\bibitem{rj01}
Romani R., Johnston S. 2001, ApJ, 557, L93

\bibitem{s97}
Shibata, S. 1997, MNRAS, 287, 262

\bibitem{s71}
Sturrock P. A., 1971, ApJ, 164, 529

\bibitem{w98}
Weatherall J. 1998, ApJ, 506, 341

\bibitem{zh00}
Zhang B.,  Harding A. K., 2000, ApJ 532, 1150

\end{thebibliography}
\end{document}